\documentclass[conference]{IEEEtran}
\IEEEoverridecommandlockouts
\usepackage{cite}
\usepackage{amsmath,amssymb,amsfonts}
\usepackage{algorithmic}
\usepackage{graphicx}
\usepackage{textcomp}
\usepackage{xcolor}
\usepackage{hyperref}
\usepackage{booktabs}   
\usepackage{multirow}   
\def\BibTeX{{\rm B\kern-.05em{\sc i\kern-.025em b}\kern-.08em
    T\kern-.1667em\lower.7ex\hbox{E}\kern-.125emX}}
\begin{document}

\title{MeloCodec: Harnessing Melodic Priors for High-Fidelity Singing Voice Representation%
\thanks{\textsuperscript{\textdagger}Corresponding authors}
}

\author{
\IEEEauthorblockN{
Yizhong Geng\IEEEauthorrefmark{1},
Wenxin Fu\IEEEauthorrefmark{1},
Kecan Mao\IEEEauthorrefmark{1},
Qifei Li\IEEEauthorrefmark{1},
Yingming Gao\IEEEauthorrefmark{1}, \\
Ruimin Wang\IEEEauthorrefmark{2},
Chunfeng Wang\IEEEauthorrefmark{2},
Hao Li\IEEEauthorrefmark{2},
Ya Li\IEEEauthorrefmark{1}\textsuperscript{\textdagger},
Wei Chen\IEEEauthorrefmark{2}\textsuperscript{\textdagger}
}
\IEEEauthorblockA{\IEEEauthorrefmark{1}Beijing University of Posts and Telecommunications, Beijing, China}
\IEEEauthorblockA{\IEEEauthorrefmark{2}Li Auto, Beijing, China}
}

\maketitle

\begin{abstract}
Neural audio codecs serve as fundamental tokenizers for LLM-based audio generation. While semantic priors are widely exploited to enhance linguistic intelligibility, the integration of explicit acoustic priors remains underexplored, limiting synthesis fidelity in frequency-sensitive domains. To address this gap, we introduce MeloCodec, a novel framework designed to effectively incorporate melodic priors—a critical form of acoustic information for singing. \textbf{To address the optimization instability typically caused by the direct fusion of such explicit priors,} we propose a \textbf{``Tokenize-then-Fuse'' paradigm} that pre-trains a discrete melodic branch to lock in structures before feature fusion. To robustly realize this paradigm, we further propose a robust two-stage training strategy, which prevents codebook collapse and ensures stable convergence. Experiments show that MeloCodec outperforms baselines in singing voice representation, improving pitch consistency and enabling controllable pitch manipulation with minimal timbre degradation. Audio samples are available at \url{https://anonymous.4open.science/api/repo/melocodec_demo-60EC/file/demo.html?v=42c197c7}.
\end{abstract}


\section{Introduction}
Neural audio codecs have fundamentally transformed audio representation learning. Early models focused on high-fidelity waveform reconstruction at extremely low bitrate as efficient alternatives to traditional DSP standards \cite{zeghidour2021soundstream, defossez2022high}. Following the success of LLMs in NLP \cite{brown2020language} and CV \cite{ramesh2021zero}, neural codecs have shifted from pure compression to discrete token modeling: vector-quantized audio codes bridge continuous waveforms and autoregressive generation, turning codecs into foundational “tokenizers” for generative AI. Systems such as VALL-E \cite{wang2023neural}, AudioLM \cite{borsos2023audiolm}, and MusicGen \cite{copet2023simple} rely on these discrete representations to capture complex acoustic dependencies, making the codec latent space quality and structure critical to generation performance.

To satisfy LLM-based generation demands, recent works actively integrate semantic priors into the quantization process. Typical systems like X-Codec \cite{ye2025codec}, SpeechTokenizer \cite{zhang2023speechtokenizer}, and BiCodec \cite{wang2025spark} fuse SSL vectors, while CosyVoice2 \cite{du2024cosyvoice} leverages ASR supervision to enhance linguistic alignment. However, prioritizing abstract semantics often neglects the intricate acoustic details essential for frequency-sensitive domains like singing voice \cite{han2025quantize, geng2025scaling}. Conversely, while approaches like FACodec \cite{ju2024naturalspeech} attempt to model acoustics via implicit latent disentanglement, they lack rigid structural guarantees, often relying on unstable multi-objective constraints that are prone to information leakage.

To overcome the limitations of semantic-only approaches and implicit disentanglement, we propose \textbf{MeloCodec}, a novel neural audio codec that explicitly integrates melodic priors into the quantization process. 
While recent emerging works have begun to explore explicit priors like Chromagrams for synthesis \cite{zhang2025vevo2, li2024accompanied, liu2025moee}, simply concatenating them within a codec's latent space introduces a critical challenge: Shortcut Learning \cite{zhao2025comelsinger}. Since the acoustic branch (processing raw waveforms) inherently contains superset information of the melody branch, the decoder tends to ignore the explicit melodic features, rendering the fusion ineffective \cite{banerjee2025neural, bous2023vasab}. Furthermore, the direct fusion of these deterministic priors with flexible neural latents causes severe optimization instability due to gradient conflicts \cite{yang2024generative, banerjee2025neural, zhao2025comelsinger, bai2025hq, sun2026muse}.

To address these challenges, we introduce \textbf{MeloCodec}, a novel framework built upon a \textbf{``Tokenize-then-Fuse'' paradigm}. By enforcing an Information Bottleneck, this paradigm tokenizes melodic priors into quantized tokens to lock in structures before feature fusion, thereby reducing timbral leakage and encouraging the acoustic stream to rely on the melodic branch. To robustly realize this, we further propose a two-stage training strategy that resolves feature conflicts, prevents codebook collapse, and ensures stable convergence where direct fusion fails. Ultimately, MeloCodec establishes a scalable blueprint for bridging interpretable physical models with generative neural networks, proving that imposing discrete bottlenecks is essential for structurally grounded LLM-based audio generation.

\begin{figure*}[t]
  \centering
  \includegraphics[width=\linewidth]{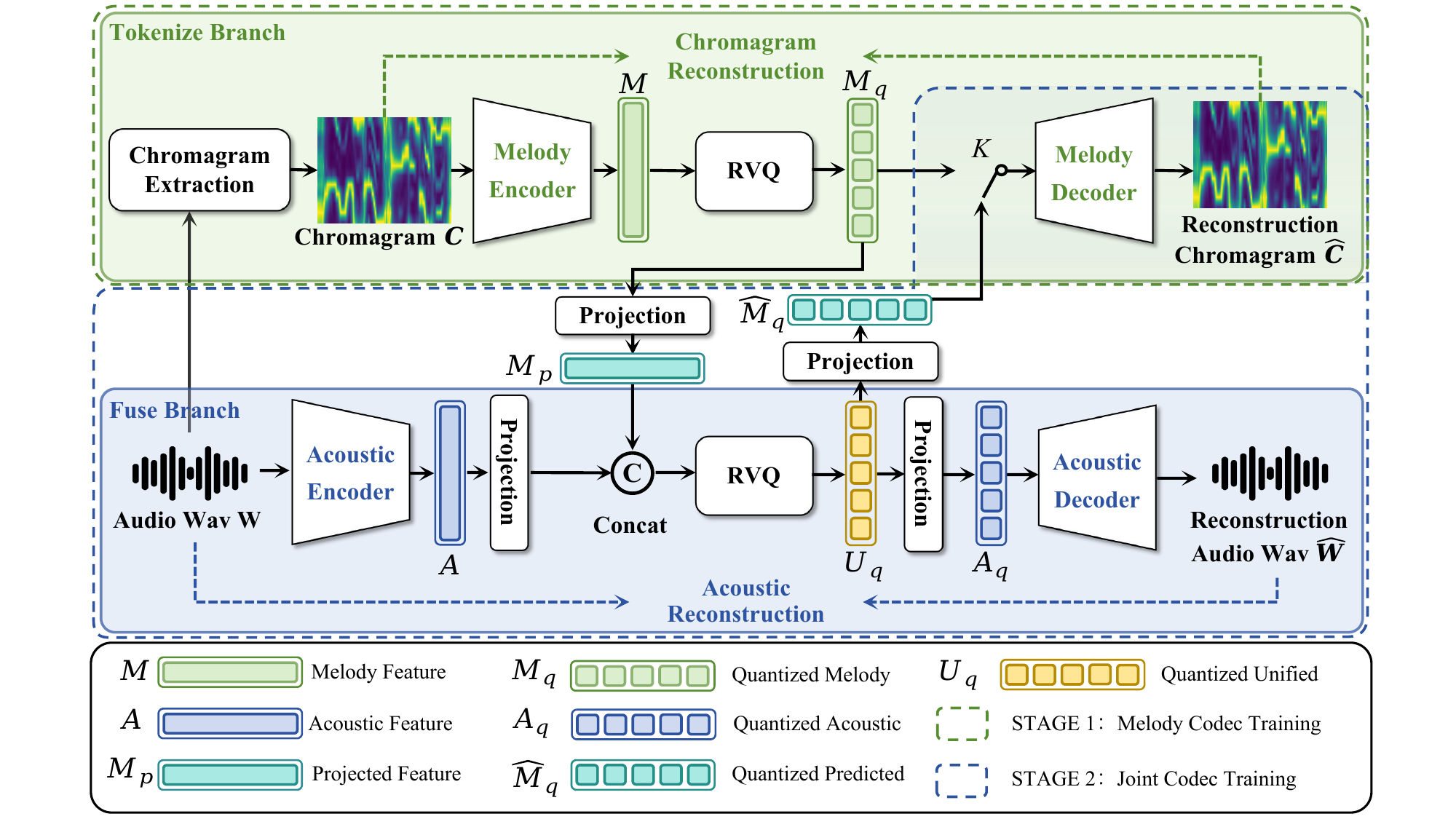}
\caption{The architecture of MeloCodec implementing the \textbf{``Tokenize-then-Fuse'' paradigm}. 
The framework consists of a \textbf{Tokenize Branch} (Green) for discrete structural modeling and a \textbf{Fuse Branch} (Blue) for acoustic reconstruction.
The switch $K$ governs the data flow across the robust \textbf{two-stage training strategy}, where dashed boxes explicitly delimit the scope of active gradient updates:
(1) In \textbf{Stage 1}, $K$ routes the quantized melody $M_q$ directly to the decoder to pre-train the bottleneck;
(2) In \textbf{Stage 2}, $K$ switches to the predicted auxiliary path ($\widehat{M}_q$), allowing gradients to update the Fuse Branch and fine-tune the Melody Decoder, while the pre-trained Melody Encoder and RVQ remains frozen.}
\label{fig:main_architecture}
\label{fig:main_architecture}
\end{figure*}

Our principal contributions are summarized as follows:
\begin{itemize}
    \item We propose \textbf{MeloCodec}, a novel codec framework that integrates explicit melodic priors into the quantization loop. Unlike previous conditioning methods, we embed melodic structures directly into the discrete latent space, establishing a unified and disentangled token interface for LLM-based generation.
    \item We introduce a \textbf{``Tokenize-then-Fuse'' paradigm} to incorporate explicit melodic priors via an Information Bottleneck, and devise a robust two-stage training strategy to address shortcut learning and optimization instability, ensuring stable convergence.
    \item We demonstrate \textbf{State-of-the-Art Performance} through evaluations on singing voice datasets, showing that MeloCodec significantly outperforms baselines in pitch consistency. Furthermore, we verify that our model enables controllable pitch manipulation with minimal timbre degradation, confirming the effectiveness of our disentangled representation.
\end{itemize}

\section{Method}

\subsection{Overview of Architecture}
\label{sec:overview}
As shown in Fig.~\ref{fig:main_architecture}, MeloCodec decouples structural modeling from texture synthesis via a \textbf{``Tokenize-then-Fuse''} paradigm. Given waveform $\mathbf{x}$, the framework comprises two interacting streams: 1) The \textbf{Tokenize Branch} extracts melodic priors (chromagrams) and quantizes them into tokens $\mathbf{M}_q$ via RVQ, creating an information bottleneck to filter timbral noise; 2) The \textbf{Fuse Branch} encodes acoustic features $\mathbf{A}$ and integrates the pre-calculated $\mathbf{M}_q$ for high-fidelity reconstruction. To address the severe optimization instability of direct fusion, we employ a two-stage training strategy.

\subsection{Tokenize Branch: Discrete Melodic Modeling} \label{sec:melody_stream} 
The \textbf{Tokenize Branch} leverages chromagrams to capture a speaker-invariant "melodic priors" \cite{zhang2025anyaccomp}. By explicitly isolating structural progressions from spectral details, this branch creates a disentangled melodic reference for the system.

Given the input audio waveform $\mathbf{x} \in \mathbb{R}^{T}$, we first extract the chromagram $\mathbf{C} \in \mathbb{R}^{F \times N}$ (where $F=12$ for pitch classes and $N$ for time frames) using Short-Time Fourier Transform (STFT) followed by pitch-class projection: $\mathbf{C} = \Phi(\mathbf{x})$, where $\Phi(\cdot)$ acts as a physical filter to discard envelope information. The chromagram is then encoded by a convolutional melody encoder $\mathcal{E}_{mel}$ into continuous latent features $\mathbf{M}$:
\begin{equation}
    \mathbf{M} = \mathcal{E}_{mel}(\mathbf{C}), \quad \mathbf{M} \in \mathbb{R}^{d \times N'}
\end{equation}
where $d$ is the hidden channel dimension and $N'$ is the downsampled temporal resolution.

To discretize $\mathbf{M}$ into compact tokens, we apply RVQ with $N_q$ levels and codebook $\mathcal{Z}$, accumulating residuals for a coarse-to-fine representation:
\begin{equation}
    \mathbf{M}_q = \sum_{k=1}^{N_q} \mathbf{z}_k, \quad \text{where } \mathbf{z}_k = \text{Quantize}_k\left(\mathbf{M} - \sum_{j=1}^{k-1} \mathbf{z}_j\right)
\end{equation}
This information bottleneck filters redundant noise, enforcing the latent space to capture essential structural semantics. A symmetric melody decoder $\mathcal{D}_{mel}$ reconstructs $\widehat{\mathbf{C}} = \mathcal{D}_{mel}(\mathbf{M}_q)$. We pre-train this branch independently using the VQ-VAE objective:
\begin{equation}
    \mathcal{L}_{stage1} = \| \mathbf{C} - \widehat{\mathbf{C}} \|_2^2 + \| \text{sg}[\mathcal{E}_{mel}(\mathbf{C})] - \mathbf{M}_q \|_2^2
\end{equation}
where $\text{sg}[\cdot]$ denotes the stop-gradient operator. This pre-training explicitly locks in structures, establishing a robust melodic "skeleton" that prevents codebook collapse in later joint optimization.

\subsection{Fuse Branch: Heterogeneous Feature Fusion}
\label{sec:acoustic_stream}
The \textbf{Fuse Branch} integrates discrete priors ($\mathbf{M}_q$) with acoustic features. Crucially, fusing \textit{quantized} tokens imposes an information bottleneck, preventing the shortcut learning and optimization instability inherent in continuous feature fusion.

The raw waveform $\mathbf{x}$ is processed by a convolutional acoustic encoder $\mathcal{E}_{ac}$ to yield dense features $\mathbf{A}$. Concurrently, the quantized melody tokens $\mathbf{M}_q$ from the Tokenize Branch are aligned to the acoustic space via a linear projection $\psi(\cdot)$:
\begin{equation}
    \mathbf{A} = \mathcal{E}_{ac}(\mathbf{x}), \quad \mathbf{M}_p = \psi(\mathbf{M}_q)
\end{equation}
where $\mathbf{A}, \mathbf{M}_p \in \mathbb{R}^{d \times N'}$, with $d$ as the hidden channel dimension and $N'$ as the downsampled temporal resolution matching the melodic stream. Fusing $\mathbf{M}_q$ rather than continuous $\mathbf{M}$ imposes a discrete information bottleneck, preventing shortcut learning where the decoder bypasses structural priors by relying on residual acoustic details.

The fused representation $\mathbf{U}$ is obtained via concatenation and then discretized into unified tokens $\mathbf{U}_q$ via a RVQ:
\begin{equation}
    \mathbf{U} = \text{Concat}(\mathbf{A}, \mathbf{M}_p)
\end{equation}
To enforce melodic integrity in $\mathbf{U}_q$, an auxiliary projection head $\phi(\cdot)$ predicts $\widehat{\mathbf{M}}_q = \phi(\mathbf{U}_q)$, which reconstructs the chromagram through the melody decoder. The auxiliary supervision loss is defined as:
\begin{equation}
    \mathcal{L}_{aux} = \| \mathbf{C} - \mathcal{D}_{mel}(\widehat{\mathbf{M}}_q) \|_2^2
\end{equation}
This supervision backpropagates gradients from the pre-trained decoder, embedding acoustic information into the unified representation without dominating acoustic textures.

\subsection{Training Strategy: Robust Two-Stage Optimization for ``Tokenize-then-Fuse''}
\label{sec:training}
To robustly realize the \textbf{``Tokenize-then-Fuse'' paradigm} and resolve the severe optimization instability arising from direct fusion between deterministic priors and flexible neural latents, we employ a robust two-stage training strategy.
This strategy first isolates the Tokenize Branch to lock in structures for stability, and then jointly optimizes the Fuse Branch with anchored discrete priors, ensuring stable convergence where direct fusion fails.

In \textbf{Stage 1 (Melody Codec Pre-training)}, only the Tokenize Branch is active: the melody encoder $\mathcal{E}_{mel}$, RVQ, and decoder $\mathcal{D}_{mel}$ are fully trained using $\mathcal{L}_{stage1}$ (Eq.~3) to build a compact codebook of pitch transitions. This phase effectively locks in structures, keeping them free from acoustic interference.

In \textbf{Stage 2 (Joint Codec Training)}, we employ a partial freezing strategy: while $\mathcal{E}_{mel}$ and its RVQ remain frozen to anchor discrete priors, the melody decoder $\mathcal{D}_{mel}$ is fine-tuned. Empirically, we observe that a frozen decoder imposes overly rigid manifold constraints, causing optimization conflicts. Unfreezing $\mathcal{D}_{mel}$ allows it to accommodate distribution shifts in the unified tokens. This strategy effectively prevents codebook collapse and guarantees the stable convergence of the joint optimization.
The Fuse Branch (Acoustic Stream) is optimized end-to-end. Specifically, $\mathcal{L}_{rec}$ comprises L1 waveform loss and multi-scale STFT loss. We employ a MSD for $\mathcal{L}_{adv}$, trained via a hinge loss objective with feature matching regularization. The total objective is:
\begin{equation}
\mathcal{L}_{stage2} = \mathcal{L}_{rec}(\mathbf{x}, \widehat{\mathbf{x}}) + \lambda_{adv} \mathcal{L}_{adv} + \lambda_{aux} \mathcal{L}_{aux}
\end{equation}
This setup ensures robust supervision flow with minimal timbre degradation.

\begin{table*}[t]
\centering
\caption{Objective performance comparison with state-of-the-art baselines on the OpenCpop test set. To ensure fair comparison, we evaluate models under two distinct bitrate settings: \textbf{high bitrate ($\approx$ 6.0 kbps)} and \textbf{low bitrate ($\approx$ 1.5 kbps)}. The results highlight MeloCodec's robustness in preserving melodic structure even under strictly constrained bandwidths.}
\label{tab:main_results}
\resizebox{\textwidth}{!}{%
\begin{tabular}{l|c|ccc|cc|cc|c}
\toprule
\multirow{2}{*}{\textbf{Model}} & \textbf{Bitrate} & \multicolumn{3}{c|}{\textbf{Architecture}} & \multicolumn{2}{c|}{\textbf{General Audio Quality}} & \multicolumn{2}{c|}{\textbf{Melody \& Pitch}} & \textbf{Timbre} \\
\cmidrule(lr){3-5} \cmidrule(lr){6-7} \cmidrule(lr){8-9} \cmidrule(lr){10-10}
 & (kbps) & \begin{tabular}[c]{@{}c@{}}Codebook\\ Size\end{tabular} & $N_q$ & \begin{tabular}[c]{@{}c@{}}Token Rate\\ (Hz)\end{tabular} & ViSQOL $\uparrow$ & STOI $\uparrow$ & F0-RMSE $\downarrow$ & Chroma-Sim $\uparrow$ & SPK-SIM $\uparrow$ \\
\midrule
\multicolumn{10}{c}{\textit{\textbf{high bitrate Setting ($\approx$6.0 kbps)}}} \\
\midrule
Encodec \cite{defossez2022high}      & 6.00 & 1024 & 8 & 75 & 3.92 & 0.82 & 1.92 & 0.93 & 0.97 \\
DAC \cite{kumar2023high}             & 6.00 & 1024 & 8 & 75 & 3.97 & 0.83 & 1.23 & 0.96 & 0.98 \\
X-Codec \cite{ye2025codec}           & 6.00 & 1024 & 8 & 75 & 3.34 & 0.81 & 1.35 & 0.95 & 0.96 \\
\textbf{MeloCodec (Ours)}            & 6.00 & 1024 & 8 & 75 & \textbf{4.08} & \textbf{0.85} & \textbf{0.96} & \textbf{0.97} & \textbf{0.99} \\
\midrule
\multicolumn{10}{c}{\textit{\textbf{low bitrate Setting ($\approx$1.5 kbps)}}} \\
\midrule
Encodec \cite{defossez2022high}      & 1.50 & 1024 & 2 & 75 & 2.81 & 0.68 & 4.87 & 0.81 & 0.88 \\
DAC \cite{kumar2023high}             & 1.50 & 1024 & 2 & 75 & 3.05 & 0.71 & 4.12 & 0.84 & 0.91 \\
X-Codec \cite{ye2025codec}           & 1.50 & 1024 & 2 & 75 & 2.74 & 0.66 & 4.68 & 0.83 & 0.87 \\
MuCodec \cite{xu2024mucodec}         & 1.33 & 10000 & 4 & 25 & 2.62 & 0.57 & 2.68 & 0.92 & 0.80 \\
\textbf{MeloCodec (Ours)}            & 1.50 & 1024 & 2 & 75 & \textbf{3.38} & \textbf{0.78} & \textbf{1.64} & \textbf{0.95} & \textbf{0.94} \\
\bottomrule
\end{tabular}%
}
\end{table*}

\section{Experiments}
\subsection{Experimental Setup}

\textbf{Datasets and Baselines.} Models were trained on an internal corpus of 5,500 hours of Mandarin singing voice (44.1kHz) and evaluated on the held-out Opencpop dataset \cite{wang2022opencpop}. To isolate architectural contributions, we compared MeloCodec against EnCodec \cite{defossez2022high}, DAC \cite{kumar2023high}, and X-Codec \cite{ye2025codec}, all fine-tuned on our internal data. We also include MuCodec \cite{xu2024mucodec} (official weights) as a zero-shot reference for low-bitrate music, as its training code is unavailable.

\textbf{Implementation Details.} We use a codebook size of 1024. Performance is evaluated under \textbf{high bitrate} ($N_q=8, \approx$6.0 kbps) and \textbf{low bitrate} ($N_q=2, \approx$1.5 kbps) settings. Crucially, melodic tokens are priors only; reported bitrates are calculated exclusively on transmitted unified tokens $\mathbf{U}_q$. Training employed AdamW and standard EMA codebook updates on NVIDIA H800 GPUs. Inference uses a sliding window (50\% overlap) to ensure continuity.

\textbf{Metrics.} Audio quality is measured via ViSQOL \cite{hines2015visqol} and STOI \cite{taal2011algorithm}; pitch accuracy via F0-RMSE and Chroma-SIM; and timbre via SPK-SIM (ResNet-34) \cite{koonce2021resnet}. Subjective quality was assessed via a MUSHRA test \cite{schoeffler2015towards} (15 experts, 30 samples). Controllability is quantified by Target F0-Corr under pitch shifting ($\pm 2, \pm 4$ semitones).

\subsection{Reconstruction Performance}
\label{sec:reconstruction}

\textbf{Objective Evaluation.} 
Table \ref{tab:main_results} shows that in the high bitrate setting, MeloCodec outperforms the strong baseline DAC \cite{kumar2023high}, confirming that integrating melodic priors enhances fidelity without interference. 

The advantages are most typical in the low bitrate setting. As bandwidth tightens ($N_q$ reduced to 2), traditional codecs degrade sharply in pitch consistency (e.g., Encodec's F0-RMSE nearly triples), suggesting standard RVQ neglects essential acoustic structural information. 
In contrast, MeloCodec remains robust (F0-RMSE 1.64). Furthermore, unlike MuCodec \cite{xu2024mucodec}, which preserves melody but suffers in general quality (ViSQOL 2.62) due to low frame rates, MeloCodec effectively bridges this gap, delivering superior melodic integrity with minimal timbre degradation. 

\begin{figure}[t]
  \centering
  \includegraphics[width=0.9\linewidth]{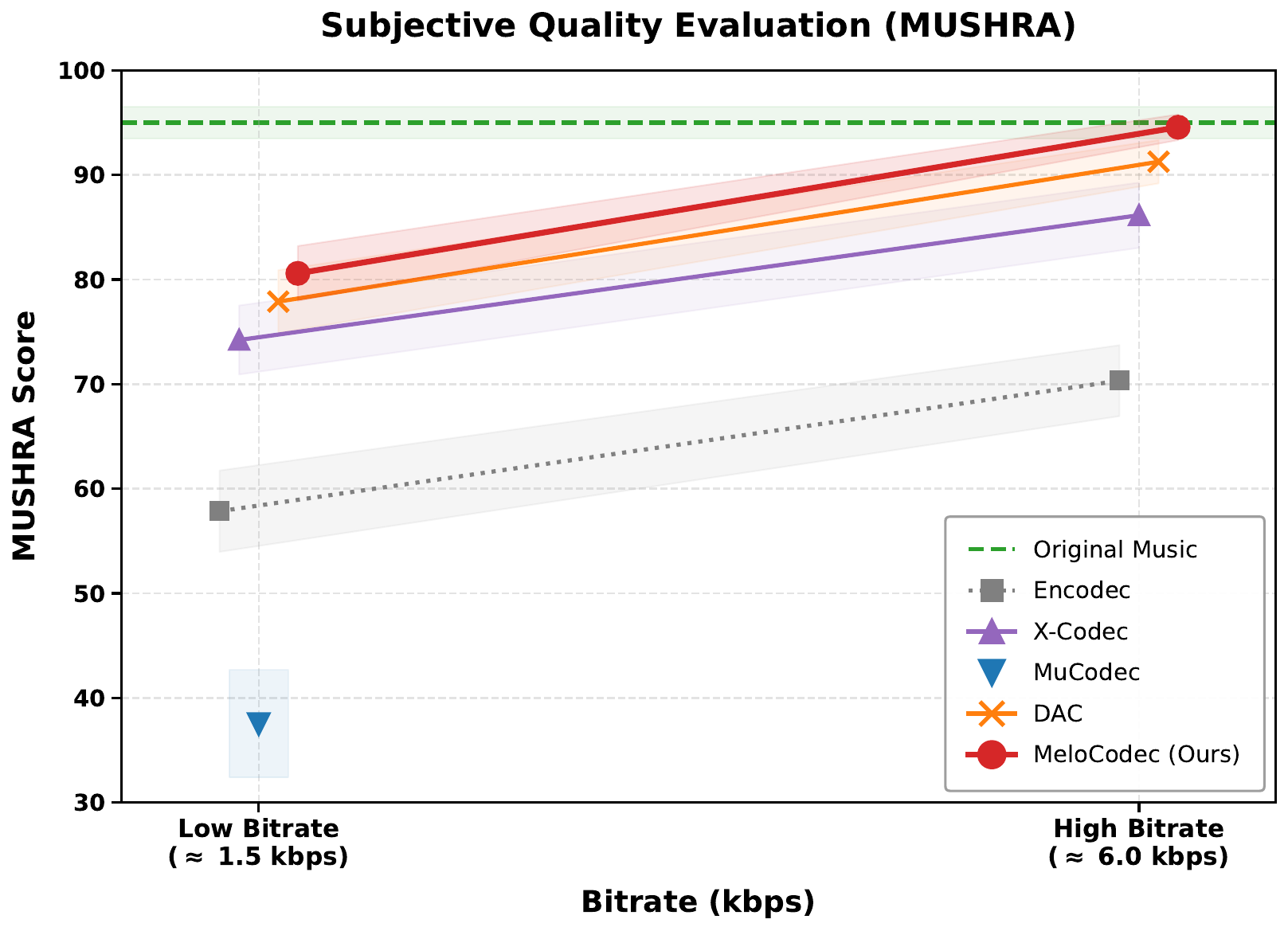}
  \caption{Subjective MUSHRA evaluation results comparing MeloCodec with baselines across varying bitrates. The shaded areas represent the 95\% confidence intervals.}
  \label{fig:mushra_score}
\end{figure}

\textbf{Subjective Evaluation.} 
To corroborate the objective metrics, we conducted a MUSHRA (Multi-Stimulus Test with Hidden Reference and Anchor) listening test to evaluate perceptual quality. As illustrated in Figure \ref{fig:mushra_score}, the subjective results align consistently with the objective data. At high bitrate, MeloCodec achieves the highest mean score, closely approaching the reference recording. Crucially, in the low-bitrate scenario, MeloCodec maintains robust performance superiority. While listeners reported noticeable artifacts and pitch instability in samples generated by baselines such as X-Codec and Encodec, MeloCodec retained clear melodic progressions and natural timbre. This underscores the effectiveness of our auxiliary melody supervision, which acts as a stable melodic prior to prevent codebook collapse under extreme compression rates.

\subsection{Melodic Integrity and Pitch Analysis}\label{sec:pitch_analysis}

To validate the model's capability in preserving and manipulating musical structures, we conduct a focused analysis on both reconstruction stability and latent disentanglement.

\begin{figure}[t]
  \includegraphics[width=\linewidth]{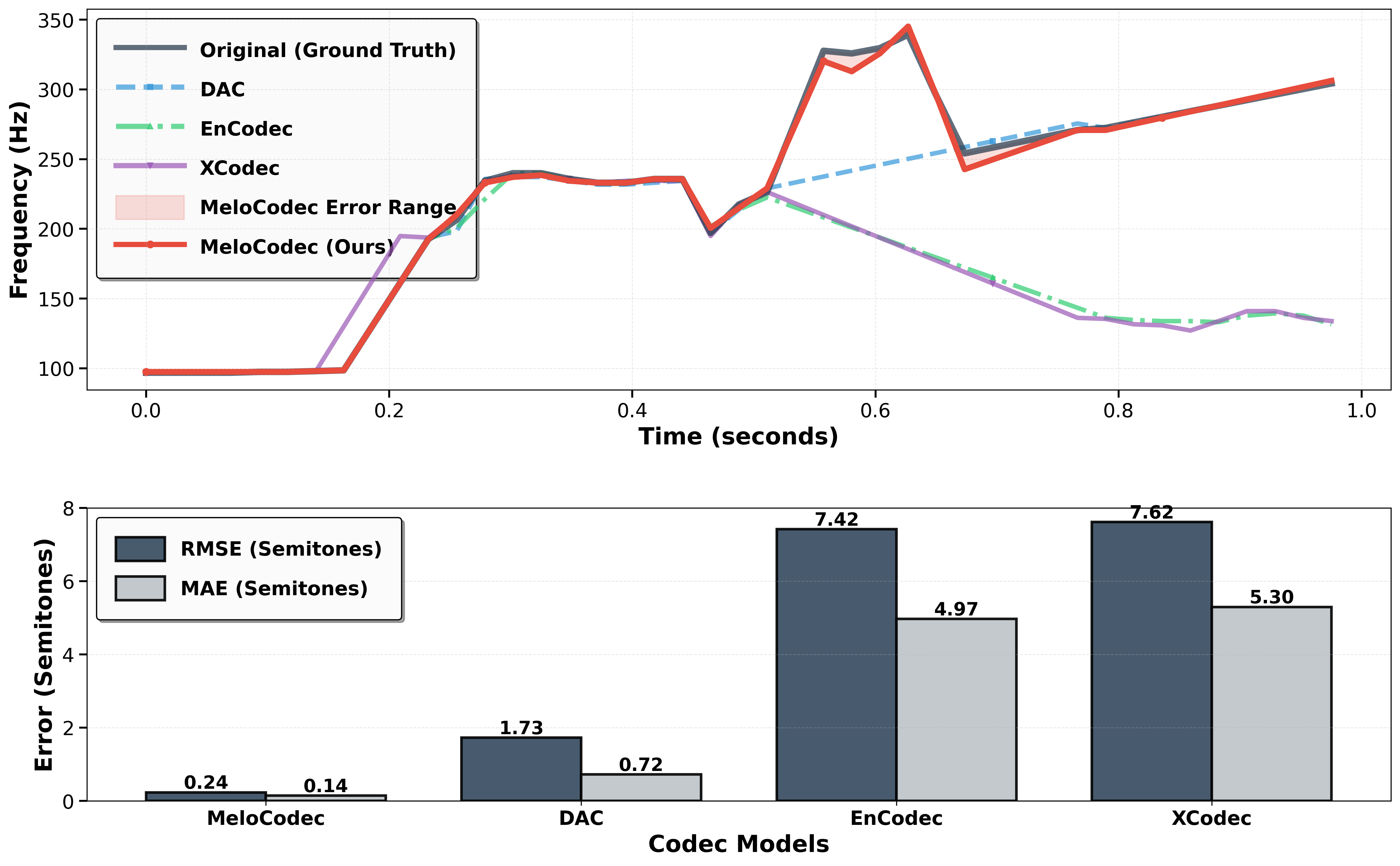}
    \caption{Visual comparison of reconstructed pitch contours (F0). 
The upper panel shows F0 trajectories with the shaded area indicating MeloCodec's deviation. 
The lower panel presents error metrics RMSE and MAE.}
\label{fig:pitch_performance}
\end{figure}

\textbf{Reconstruction Stability.} As detailed in Table \ref{tab:main_results}, MeloCodec consistently surpasses baselines in pitch accuracy. The advantage is most critical in the low bitrate setting ($\approx$1.5 kbps). MeloCodec remains robust (F0-RMSE 1.64), acting as a structural anchor. This is visually corroborated in Figure \ref{fig:pitch_performance}, where our model captures subtle vibratos without the frequency jitter and octave errors observed in baseline reconstructions.

\textbf{Disentanglement \& Controllability.} We evaluate controllability by shifting melodic priors ($\pm 2/4$ semitones) while keeping acoustic features fixed. As shown in Table \ref{tab:pitch_control}, \textbf{MeloCodec} maintains robust contour correlation (0.70--0.72), confirming effective melodic drive. However, the elevated RMSE (3.85--4.50) reveals a trade-off: while the contour shifts, absolute pitch precision remains partially constrained by the fixed acoustic branch. In contrast, Direct Fusion fails completely (RMSE 6.80), confirming its vulnerability to shortcut learning.

\begin{table*}[t]
\centering
\caption{Ablation study of MeloCodec training strategies on the OpenCpop test set. We compare different fusion and supervision methods using Chromagram features. The results highlight that the proposed ``Tokenize-then-Fuse'' paradigm (Ours) is essential for stability and performance, especially at low bitrate.}
\label{tab:ablation_study}
\resizebox{\textwidth}{!}{%
\begin{tabular}{l|c|ccc|cc|cc|c}
\toprule
\multirow{2}{*}{\textbf{Method}} & \textbf{Bitrate} & \multicolumn{3}{c|}{\textbf{Architecture}} & \multicolumn{2}{c|}{\textbf{General Audio Quality}} & \multicolumn{2}{c|}{\textbf{Melody \& Pitch}} & \textbf{Timbre} \\
\cmidrule(lr){3-5} \cmidrule(lr){6-7} \cmidrule(lr){8-9} \cmidrule(lr){10-10}
 & (kbps) & \begin{tabular}[c]{@{}c@{}}Codebook\\ Size\end{tabular} & $N_q$ & \begin{tabular}[c]{@{}c@{}}Token Rate\\ (Hz)\end{tabular} & ViSQOL $\uparrow$ & STOI $\uparrow$ & F0-RMSE $\downarrow$ & Chroma-Sim $\uparrow$ & SPK-SIM $\uparrow$ \\
\midrule

\multicolumn{10}{c}{\textit{\textbf{high bitrate Setting ($\approx$6.0 kbps)}}} \\
\midrule
Acoustic-only & 6.00 & 1024 & 8 & 75 & 3.97 & 0.83 & 1.23 & 0.96 & 0.98 \\
+ Auxiliary Loss & 6.00 & 1024 & 8 & 75 & 3.98 & 0.84 & 1.15 & 0.96 & 0.98 \\
Direct Fusion & 6.00 & 1024 & 8 & 75 & 3.52 & 0.79 & 2.30 & 0.93 & 0.96 \\
\textbf{Ours (Tokenize-then-Fuse)} & 6.00 & 1024 & 8 & 75 & \textbf{4.08} & \textbf{0.85} & \textbf{0.96} & \textbf{0.97} & \textbf{0.99} \\
\midrule

\multicolumn{10}{c}{\textit{\textbf{low bitrate Setting ($\approx$1.5 kbps)}}} \\
\midrule
Acoustic-only & 1.50 & 1024 & 2 & 75 & 3.05 & 0.71 & 4.12 & 0.84 & 0.91 \\
+ Auxiliary Loss & 1.50 & 1024 & 2 & 75 & 3.25 & 0.75 & 2.10 & 0.91 & 0.93 \\
Direct Fusion & 1.50 & 1024 & 2 & 75 & 2.34 & 0.58 & 5.95 & 0.81 & 0.85 \\
\textbf{Ours (Tokenize-then-Fuse)} & 1.50 & 1024 & 2 & 75 & \textbf{3.38} & \textbf{0.78} & \textbf{1.64} & \textbf{0.95} & \textbf{0.94} \\
\bottomrule
\end{tabular}%
}
\end{table*}

\begin{table}[t]
\centering
\caption{Evaluation of disentanglement via Pitch Shifting. Note that while absolute pitch precision (RMSE) is limited by acoustic constraints, MeloCodec maintains superior melodic contour (Corr) compared to Direct Fusion.}
\label{tab:pitch_control}
\resizebox{\columnwidth}{!}{%
\begin{tabular}{l|c|cc|cc}
\toprule
\multirow{2}{*}{\textbf{Method}} & \multirow{2}{*}{\textbf{Shift}} & \multicolumn{2}{c|}{\textbf{Pitch Control}} & \multicolumn{2}{c}{\textbf{Timbre Stability}} \\
\cmidrule(lr){3-4} \cmidrule(lr){5-6}
 & (Semi.) & Target F0-Corr $\uparrow$ & Target F0-RMSE $\downarrow$ & SPK-SIM $\uparrow$ & Degrade \% $\downarrow$ \\
\midrule
\multirow{2}{*}{Direct Fusion} & $\pm 2$ & 0.40 & 4.10 & 0.88 & 7.6\% \\
 & $\pm 4$ & 0.27 & 6.80 & 0.82 & 14.0\% \\
\midrule
\multirow{2}{*}{\textbf{MeloCodec}} & $\pm 2$ & \textbf{0.72} & \textbf{3.85} & \textbf{0.93} & \textbf{1.6\%} \\
 & $\pm 4$ & \textbf{0.70} & \textbf{4.50} & \textbf{0.91} & \textbf{3.9\%} \\
\bottomrule
\end{tabular}%
}
\end{table}

\subsection{Ablation Study: Efficacy of Training Strategies}
\label{sec:ablation}

To validate the advantage of our proposed ``Tokenize-then-Fuse'' paradigm, we conducted ablation studies comparing MeloCodec against three variants: (1) \textbf{Acoustic-only}, a standard single-stream codec similar to DAC; (2) \textbf{+ Auxiliary Loss}, single-stream codec trained with additional melody reconstruction supervision; and (3) \textbf{Direct Fusion}, a dual-stream architecture trained end-to-end without the two-stage training strategy. 
The results are summarized in Table \ref{tab:ablation_study}.

\textbf{Benefit of Melodic Supervision.}
Comparing the first two rows, adding an auxiliary melody loss yields a noticeable improvement in pitch accuracy, reducing F0-RMSE from 4.12 to 2.10 in the low-bitrate setting. This confirms that explicit supervisory signals can guide the encoder to retain more pitch information. However, the gain in general audio quality (ViSQOL) is marginal, suggesting that soft constraints via loss functions alone are insufficient to fundamentally restructure the latent space.

\textbf{Instability of Direct Fusion.}
We analyze the Direct Fusion strategy, representing the standard "One-stage" paradigm. To ensure fairness, this baseline employs learnable projection layers (MLP) and normalization to align heterogeneous modalities, avoiding simple concatenation. Despite these safeguards, the method exhibits catastrophic degradation (F0-RMSE worsens to 5.95), confirming severe optimization instability. 

Diagnostic analysis of training dynamics reveals the root cause as modality collapse driven by feature competition. We observed two failure modes:
1) \textbf{Codebook Collapse}: Codebook utilization dropped to $<5\%$ early in training, indicating the discrete bottleneck was bypassed.
2) \textbf{Optimization Divergence}: Crucially, the auxiliary chroma validation loss diverged continuously while training loss decreased. 
This confirms shortcut learning: dense acoustic gradients dominate optimization, causing the decoder to overfit acoustic textures rather than attending to rigid melodic priors. This verifies that standard architectural tuning (e.g., scaling or gating) is insufficient; rather, the proposed ``Tokenize-then-Fuse'' paradigm is structurally necessary to enforce effective melodic encoding.

\textbf{Effectiveness of ``Tokenize-then-Fuse''.}
In contrast, our strategy achieves the best overall performance. By decoupling training into two stages and fusing quantized melody tokens ($M_q$), we effectively resolve severe optimization instability. The pre-trained Melodic Stream explicitly locks in structures, allowing the Acoustic Stream to focus on texture synthesis. This is evident at low bitrate, where our method ensures stable convergence (F0-RMSE 1.64) compared to the collapse in Direct Fusion.

\section{Conclusion}
\label{sec:conclusion}

In this paper, we presented \textbf{MeloCodec}, a dual-stream neural audio codec that explicitly integrates melodic priors to achieve high-fidelity singing voice representation. By addressing the severe optimization instability inherent in the direct fusion of features, our proposed ``Tokenize-then-Fuse'' paradigm effectively locks in structures via a pre-trained discrete branch, ensuring stable convergence and preventing codebook collapse. Results demonstrate that this strategy significantly outperforms baselines in pitch consistency with minimal timbre degradation. Crucially, our framework establishes a scalable blueprint for bridging interpretable physical models with generative neural networks, proving that imposing discrete bottlenecks on deterministic priors is essential for robust feature integration. This opens new avenues for future research to extend the disentangled interface by incorporating additional explicit acoustic priors, such as rhythmic patterns and dynamic contours, paving the way for fully controllable and structurally grounded LLM-based audio generation.

\section{Acknowledgment}
The work was supported by the National Natural Science Foundation of China (NSFC) (No.\ 62271083), and the Key Project of the National Language Commission (No.\ ZDI145-81).

\bibliographystyle{IEEEbib}
\bibliography{icme2025references}

\end{document}